\documentclass[10pt]{article}
\usepackage{graphicx}
\begin{document}

\title{Is bimodal fission an indirect experimental evidence for pionic
radioactivity?}

\author{D. B. Ion$^{1}$, Reveica Ion-Mihai$^{2}$, M. L. Ion$^{2}$ and Adriana I.
Sandru$^{1}$\\
$^{1}$ \textit{National Institute for Physics and Nuclear Engineering,}\\
\textit{IFIN-HH}, Bucharest, Romania\\
$^{2}$ \textit{University of Bucharest, Department of Atomic and} \\
\textit{Nuclear Physics}, Bucharest, Romania}

\maketitle

\begin{abstract}
In this paper new predictions for the spontaneous pion emission
accompanied by fission are presented for all nuclei with 100$\leq $Z$\leq $%
108. The bimodal fission as an indirect experimental evidence for the pionic
radioactivity is demonstrated. The experimental tests of this important
connection are suggested.
\end{abstract}

The nuclear pionic radioactivity (NPIR) of a parent nucleus (A,Z), introduced in
1985 by Ion, Ivascu and R. Ion-Mihai [1-4], can be considered as an
inclusive reaction of form:

\begin{equation}
(A,Z)\rightarrow \pi +X  \label{1}
\end{equation}

\noindent where X denotes any configuration of final particles (fragments, light
neutral and charged particles, etc.) which accompany emission process. The
inclusive NPIR is in fact a sum of all exclusive nuclear reactions allowed
by the conservation laws in which a pion can be emitted by a nucleus from
its ground state. The most important exclusive reactions which give the
essential contribution to the inclusive NPIR (1) are the spontaneous pion
emission accompanied by two body fission:

\begin{equation}
_{Z}^{A}X\rightarrow \pi +_{Z_{1}}^{A_{1}}X+_{Z_{2}}^{A_{2}}X  \label{2}
\end{equation}

\noindent where $A=A_1+A_2$ and $Z=Z_1+Z_2+Z_\pi $ .

Hence, the NPIR is an extremely complex coherent reaction in which we are
dealing with a spontaneous pion emission accompanied by a rearrangement of
the parent nucleus in two or many final nuclei. Charged pions as well as
neutral pions can be emitted during two body or many body fission of parent
nucleus.

The energy liberated in an exclusive nuclear reaction (1) is given by

\begin{equation}
Q_{\pi F}=Q_{n}-m_{\pi }  \label{3}
\end{equation}

\noindent where $Q_n$ is the energy liberated in n-body spontaneous fission

\begin{equation}
Q_{n}=m(Z,A)-m(Z_{1},A_{1})-m(Z_{2},A_{2})-...-m(Z_{n},A_{n})  \label{4}
\end{equation}

In the last decades the studies of the spontaneous fission (SF) properties
of heavy nuclides in the region of Fermium contributed essentially to a real
progress in the understanding [5-9] of the sudden transition from the
asymmetric mass division observed in the SF of all the lighter actinides to
the very symmetric mass division accompanied by anomalous high total kinetic
energy (TKE) found in SF of heavier Fm isotopes. This transition was first
found [9] to occur in Fm between $^{256}$Fm and $^{258}$Fm, $^{257}$Fm being
a transition nucleus with much enhanced yields and high TKE`s for the
symmetric mass division.

Therefore, the investigations (see quotations and the results in refs.[5-9])
of the spontaneous fission (SF) properties of heavy nuclides: $^{258}$Fm, $%
^{259}$Md, $^{260}$Md,$^{258}$No, and $^{260}$Ha etc., shown that all these
nuclides are fissioning with a symmetrical division of mass. The transitions
from asymmetric to symmetric mass division for the Fm isotopes as well as
for No nuclides are shown schematically in Fig. 3. The most striking feature
is narrowness of the full width at half maximum $\Delta M$ of these mass
distributions. Moreover, the total kinetic energy (TKE) distributions
strongly deviates from the Gaussian shape characteristically found in the
fission of other actinides. When the TKE-distributions are resolved in two
Gaussian distributions, one of the constituent peaks lie in the low-energy region near
200 MeV while the second peak lie at the high-energy near 233 MeV. This
property of the SF TKE-distributions is called bimodal fission. Hence, the
low-energy-SF-mode is characterized by broad symmetrical mass distribution
and fragment energy in accord with liquid-drop systematics, whereas the
high-energy-SF-mode produces sharply symmetrical mass distributions with the
TKE of fragments approaching the $Q_{SF}$ values for the fission reaction.
Even though both SF modes are possible in the same parent nuclei, but one
generally predominates. The theoretical explanation offered by Hulet et al.
(see ref. [7]) for the bimodal fission is based on the calculations of
static deformations. Each mode is derived from the shell structure effects;
one in the parent fissioning nucleus and the other from single-particle
couplings in the fragments. Theorists have found (see Refs.[11]) new
appropriate valleys which can explain the bimodal fission. However, The explanation how each
mode can coexist and occur with near equal probability presents a challenging
problem. The authors of the paper [12], starting with the observation that
this kind of symmetric fission occur for nuclei in which the initial
formation of a cluster within the valence shells of the fissioning nucleus
yields an energy Q$_2$ higher than the threshold for the creation of a pion
on mass shell, suggested that this mode of fission could be related to a new
phase of nuclear matter. So, these remarks suggest that something new
happens in the symmetric spontaneous fission of the heavy actinides. But
this something new happens only for those parent nuclei which are
spontaneously fissioning in two steps and for which a NPIR reaction (1) with
a pion on mass shell is energetically possible even in the first
intermediate step, e.g. (see the first diagram in the right hand of Fig.
4a.):

\begin{center} $
^{258}Fm\rightarrow ^{50}Ar+^{208}Pb,\; \; Q_{2}\geq m_{\pi } $
\end{center}

\noindent followed by the second step:

\begin{center} $
^{50}Ar+^{208}Pb\rightarrow ^{132}Sn+^{126}Sn+Q $
\end{center}

\noindent leading to the symmetric final state. Hence, this something new could be the
appearance of the pion on mass shell as a third constituent of the nucleus
in which the distinction between proton and neutron disappear- due the
transition $p\Leftrightarrow n$(at equilibrium)- and for which the magic
numbers remain the same as for the usual constituents n or p. These two
steps can be represented as in the first (right hand) unitarity diagram in
Fig.4a.  So, according to
the unitarity diagram from Fig. 4a, the spontaneous fission (SF) can be
described as a bimodal two step process as follows

\begin{itemize}
\item Asymmetric fission mode (A)

\begin{itemize}
\item first step:
\begin{equation}
^{258}Fm\rightarrow ^{50}Ar+^{208}Pb  \label{7}
\end{equation}

\item second step:
\begin{equation}
^{50}Ar+^{208}Pb\rightarrow ^{126}Sn+^{132}Sn  \label{8}
\end{equation}
\end{itemize}

\item Symmetric fission mode (S):
\begin{itemize}
\item first step :
\begin{equation}
^{258}Fm\rightarrow \pi +^{50}Ar+^{208}Pb  \label{9}
\end{equation}
\item second step:

\begin{equation}
\pi +^{50}Ar+^{208}Pb\rightarrow ^{126}Sn+^{132}Sn  \label{10}
\end{equation}
\end{itemize}
\end{itemize}

In similar way, for the pionic radioactivity (Fig. 1b) we have

\begin{itemize}
\item NPIR-Asymmetric mode

\begin{itemize}
\item first step:
\begin{equation}
^{258}Fm\rightarrow ^{50}Ar+^{208}Pb  \label{11}
\end{equation}
\item second step:

\begin{equation}
^{50}Ar+^{208}Pb\rightarrow \pi +^{126}Sn+^{132}Sn  \label{12}
\end{equation}
\end{itemize}
\item NPIR-Symmetric mode

\begin{itemize}
\item first step:
\begin{equation}
^{258}Fm\rightarrow \pi +^{50}Ar+^{208}Pb  \label{13}
\end{equation}

\item second step:
\begin{equation}
\pi +^{50}Ar+^{208}Pb\rightarrow \pi +^{126}Sn+^{132}Sn  \label{14}
\end{equation}
\end{itemize}
\end{itemize}

Then, the scenario described in Figs. 4, which include in
a more general and exact form the idea of transition to a new phase of
nuclear matter is leading us to point out a possible connection between the
observed bimodal symmetric SF at the nuclides $^{258}Fm$, $^{259}Fm$, $%
^{259}Md $, $^{260}Md$, $^{258}No$, with a significant high NPIR-yields for
these nuclei. In order to illustrate this idea we give in Table 1 the NPIR yields,
predicted by the fission-like model [see Ref. [13]]:

\begin{equation}
\frac{\Gamma _{\pi }}{\Gamma _{SF}}=\left[ \frac{T_{SF}}{T_{SF}^{C}}\right]
^{\delta _{\alpha }(A,Z)}  \label{15}
\end{equation}
\noindent where

\begin{equation}
\delta _{\alpha }(A,Z)=\frac{\Delta \theta _{\alpha }}{\theta -5}  \label{16}
\end{equation}
\noindent and

\begin{eqnarray}
\theta = Z^2/A-37.5\\
\Delta \theta _{\alpha }=\frac{m_{\pi }}{\gamma A^{2/3}}\frac{\alpha
-(1-\alpha )E_{C}^{0}/E_{S}^{0}}{1-(1-\alpha )m_{\pi }/E_{S}^{0}} \\
E_{C}^{0}=0.7053Z^{2}A^{-1/3}(MeV) \nonumber \\
E_{S}^{0}=17.80A^{2/3}(MeV) \nonumber
\end{eqnarray}
\noindent where for even-even parent nuclei we have
\begin{center} $
T_{SF}^{C}=10^{-10,95}(yr) $
\end{center}
\noindent and for A-odd parent nuclei

\begin{center} $
T_{SF}^{C}=10^{-8,39}(yr) $
\end{center}

Here, as in Ref. [13] we also adopt the notions of \textit{SF-nuclide}, $\pi F$%
\textit{-nuclide}, and \textit{T-transition nuclide}, as being those parent nuclei
characterized by :$\frac{\Gamma _{\pi ^{0,\pm }}}{\Gamma _{SF}}<1$ , $\frac{%
\Gamma _{\pi ^{0,\pm }}}{\Gamma _{SF}}>1$ and yield $\Gamma _{\pi ^{0,\pm
}}=\Gamma _{SF}$, respectively, with their characteristic features presented
in Table 2.

Therefore, the results presented in Table 1 can now be summarized as follows
[in parenthesis we give the predicted pionic yield for $\pi ^{0}$]:
\begin{itemize}
\item The nuclides $^{242}Fm$ ($3.3\cdot 10^{-2}$) are expected to be $\pi F$-nuclides, if
our predictions with $\alpha =1$ underestimate the pionic yield determined with Eqs.
(13)-(14). Then, a new transition nucleus can be predicted to be in the
region of $Fm$ with $A=240\div242$. The experiments with parent nuclei in this $Fm$
region are needed in order to clarify these predictions.
\item The nuclides $^A Fm$ with $244\leq A\leq 256$ are all $SF$ nuclides
since all are have asymmetric mass distribution of fission fragments (see Fig. 3).
\item The problem of the
nucleus $^{257}Fm$ ($1.4\cdot 10^{-14}$) as transition nuclide can also be supported
not only by ''bimodal'' experimental observation but also by some
experimental high pionic yields inferred in Refs.[15,16] from the
experimental data of Wild et al.,[17].
\item  The best $\pi$-emitter with symmetric
mode of fission in this region is expected to be$^{258}Fm$ (0.95). If our
interpretation of the bimodal fission via the saturation of the unitarity
limits shown in Fig. 4a,b is correct we expect that pionic yield of $^{258}Fm$
to be higher than 1. The nuclei $^{259}Fm$ ($5.1\cdot 10^{-2}$) and
$^{260}Fm$ ($5.6\cdot 10^{-2}$) are also interesting from experimental point of view
since both can be $\pi F$-nuclides.

\item The nuclides $No$ with $A=242\div250$ are expected to be all $\pi F$-nuclides
(inferred from Fig. 4 in Ref. [3]), while the $No$-isotopes with $A\geq252$ are all
$SF$-nuclides (see Table 1). High pionic yields are obtained for $^{250}No$
(6.67) as well as for $^{258}No$ ($7.1\cdot 10^{-2}$) and $^{262}No$ ($1.2 10^{-2}$).

\item The nuclides $Lr$ and $Rf$ are all predicted to be $SF$-nuclides (see Table 1).

\item The nuclides $^{258}Sq$ ($1.3\cdot 10^7$), $^{259}Sg$ ($6.1\cdot 10^6$), $^{260}Sg$
($1.2\cdot 10^8$), $^{261}Sg$ ($1.3\cdot 10^11$), $^{263}Sg$ ($5.2\cdot 10^27$) as well
as $^{264}Hs$ ($1.2\cdot 10^2$), all are predicted to be $\pi F$-nuclides.
\end{itemize}

We note of course that all these estimations must be taken as orientating for the
future experimental investigations where the $\pi F$-methods are expected to play
an important role.
Therefore, dedicated $\pi F$-experiments using nuclei with theoretically
predicted (see Table 1) high pionic yields are desired.

For sake of completeness we recall here the results [3,14] regarding the competition
between the pionic radioactivity and the spontaneous fission of the
SHE-nuclides. Indeed, using the same fission-like model, it was shown that
most of the SHE-nuclei lie in the region where the pionic fissility
parameters attain their limiting values $X_{\pi F}=1$. Hence, the SHE-region is
characterized by the absence of a classical barrier toward spontaneous pion
emission. Consequently, both decay modes $\pi$-fission and the spontaneous
fission of SHE nuclides essentially will be determined only by shell
effects. Then, it was seen that the usual SHE-island of stability around the
double magic nucleus $^{298}[114]$ can be destroyed due to the dominant pionic
radioactivity of the SHE-nuclei from this region. Therefore, we believe that
in future searches for transfermium as well as for SHE-nuclei, the $\pi$-fission
as detection method [18] can play an essential role in the discovery of this
new phase of nuclear matter produced by the presence of pions on mass shell
in the nuclear medium.
This paper is dedicated to Professor Marin Ivascu at
his 70th anniversary for his continuous effort to the introduction and
development of this new scientific frontier domain of physics as is the
nuclear pionic radioactivity.

\pagebreak \enlargethispage{3cm} \thispagestyle{empty} \vspace*{-20mm}
\noindent \textbf{Table 1:} Theoretical predictions of the $\frac{\Gamma _{\pi ^{0,\pm }}}{%
\Gamma _{SF}}$ yields for heavy parent nuclei with Z between Z=100 and
Z=108, obtained by using Eqs. (13)-(14). \nopagebreak \noindent \hspace*{-10mm}$
\begin{array}{|l|l|l|l|l|l|l|l|} \hline
& Z & A & T_{SF}(yr) & T_{SF}/T_{SF}^{C} & \Gamma _{\pi ^{-}}/\Gamma
_{SF} & \Gamma _{\pi ^{0}}/\Gamma _{SF} & \Gamma _{\pi +}/\Gamma _{SF}
\\ \hline
Fm & 100 & 242 & 2,535\cdot 10^{-11} & 2,259\cdot 10^{+00} & 3,55\cdot 10^{-02} & 3,30\cdot 10^{-02} & 2,43\cdot 10^{-02} \\
& 100 & 244 & 1,046\cdot 10^{-10} & 9,320\cdot 10^{+00 }& 8,59\cdot 10^{-04} & 7,35\cdot 10^{-04} & 3,85\cdot 10^{-04} \\
& 100 & 245 & 1,255\cdot 10^{-04} & 3,080\cdot 10^{+04} & 1,77\cdot 10^{-13} & 9,32\cdot 10^{-14} & 6,33\cdot 10^{-15} \\
& 100 & 246 & 4,753\cdot 10^{-07} & 4,236\cdot 10^{+04} & 1,17\cdot 10^{-12} & 6,39\cdot 10^{-13} & 5,16\cdot 10^{-14} \\
& 100 & 248 & 1,141\cdot 10^{-03} & 1,017\cdot 10^{+08} & 3,59\cdot 10^{-18} & 1,48\cdot 10^{-18} & 3,74\cdot 10^{-20} \\
& 100 & 250 & 8,000\cdot 10^{-01} & 7,130\cdot 10^{+10} & 3,28\cdot 10^{-21} & 1,16\cdot 10^{-21} & 1,54\cdot 10^{-23} \\
& 100 & 252 & 1,250\cdot 10^{+02} & 1,114\cdot 10^{+13} & 1,84\cdot 10^{-22} & 6,12\cdot 10^{-23} & 6,26\cdot 10^{-25} \\
& 100 & 254 & 6,242\cdot 10^{-01} & 5,563\cdot 10^{+10} & 9,36\cdot 10^{-17} & 4,15\cdot 10^{-17} & 1,41\cdot 10^{-18} \\
& 100 & 256 & 3,308\cdot 10^{-04} & 2,948\cdot 10^{+07} & 8,08\cdot 10^{-11} & 4,84\cdot 10^{-11} & 5,77\cdot 10^{-12} \\
& 100 & 257 & 1,310\cdot 10^{+02} & 3,216\cdot 10^{+10} & 2,75\cdot 10^{-14} & 1,38\cdot 10^{-14} & 7,93\cdot 10^{-16} \\
& 100 & 258 & 1,172\cdot 10^{-11} & 1,045\cdot 10^{+00} & 9,47\cdot 10^{-01} & 9,46\cdot 10^{-01} & 9,41\cdot 10^{-01} \\
& 100 & 259 & 4,753\cdot 10^{-08} & 1,167\cdot 10^{+01} & 5,44\cdot 10^{-02} & 5,10\cdot 10^{-02} & 3,91\cdot 10^{-02} \\
& 100 & 260 & 1,268\cdot 10^{-10} & 1,130\cdot 10^{+01} & 6,33\cdot 10^{-02} & 5,95\cdot 10^{-02} & 4,62\cdot 10^{-02} \\ \hline
Md & 101 & 245 & 2,852\cdot 10^{-11} & 7,001\cdot 10^{-03} & 8,66\cdot 10^{+11} & 1,58\cdot 10^{+12} & 1,96\cdot 10^{+13} \\
& 101 & 247 & 6,338\cdot 10^{-09} & 1,556\cdot 10^{+00} & 1,73\cdot 10^{-01} & 1,67\cdot 10^{-01 }& 1,42\cdot 10^{-01} \\
& 101 & 255 & 3,422\cdot 10^{-02} & 8,401\cdot 10^{+06} & 1,20\cdot 10^{-13} & 6,27\cdot 10^{-14} & 4,12\cdot 10^{-01} \\
& 101 & 257 & 6,297\cdot 10^{-02} & 1,546\cdot 10^{+07} & 1,37\cdot 10^{-12} & 7,52\cdot 10^{-13} & 6,16\cdot 10^{-14} \\
& 101 & 259 & 1,848\cdot 10^{-04} & 4,536\cdot 10^{+04} & 1,28\cdot 10^{-07} & 9,04\cdot 10^{-08} & 2,11\cdot 10^{-08} \\ \hline
No & 102 & 250 & 7,922\cdot 10^{-12} & 7,061\cdot 10^{-01} & 6,41\cdot 10^{+00} & 6,67\cdot 10^{+00} & 7,91\cdot 10^{+00} \\
& 102 & 251 & 3,169\cdot 10^{-07} & 7,779\cdot 10^{+01} & 3,33\cdot 10^{-09} & 2,17\cdot 10^{-09 }& 3,63\cdot 10^{-10} \\
& 102 & 252 & 2,852\cdot 10^{-07} & 2,542\cdot 10^{+04} & 9,39\cdot 10^{-18} & 3,96\cdot 10^{-18} & 1,09\cdot 10^{-19} \\
& 102 & 254 & 9,126\cdot 10^{-04} & 8,134\cdot 10^{+07 }& 1,02\cdot 10^{-24} & 3,02\cdot 10^{-25} & 1,92\cdot 10^{-27} \\
& 102 & 256 & 1,711\cdot 10^{-05} & 1,525\cdot 10^{+06} & 3,58\cdot 10^{-16} & 1,64\cdot 10^{-16} & 6,32\cdot 10^{-18 }\\
& 102 & 257 & 5,324\cdot 10^{-05} & 1,307\cdot 10^{+04} & 3,53\cdot 10^{-10} & 2,19\cdot 10^{-10} & 2,98\cdot 10^{-11} \\
& 102 & 258 & 3,803\cdot 10^{-11} & 3,389\cdot 10^{+00} & 7,47\cdot 10^{-02} & 7,06\cdot 10^{-02} & 5,56\cdot 10^{-02} \\
& 102 & 259 & 1,141\cdot 10^{-03} & 2,800\cdot 10^{+05 }& 1,68\cdot 10^{-11} & 9,75\cdot 10^{-12} & 1,00\cdot 10^{-12} \\
& 102 & 260 & 3,359\cdot 10^{-09} & 2,994\cdot 10^{+02} & 2,62\cdot 10^{-05} & 2,08\cdot 10^{-05} & 7,91\cdot 10^{-06} \\
& 102 & 262 & 1,584\cdot 10^{-10} & 1,412\cdot 10^{+01} & 1,30\cdot 10^{-02} & 1,18\cdot 10^{-02} & 7,96\cdot 10^{-03} \\ \hline
Lr & 103 & 253 & 4,183\cdot 10^{-06} & 1,027\cdot 10^{+03} & 1,39\cdot 10^{-25} & 3,97\cdot 10^{-26} & 2,10\cdot 10^{-28} \\
& 103 & 255 & 6,854\cdot 10^{-04} & 1,680\cdot 10^{+05} & 7,00\cdot 10^{-28} & 1,77\cdot 10^{-28 }& 5,77\cdot 10^{-31} \\
& 103 & 257 & 6,274\cdot 10^{-05} & 1,540\cdot 10^{+04} & 1,25\cdot 10^{-16} & 5,59\cdot 10^{-17} & 1,95\cdot 10^{-18} \\
& 103 & 259 & 9,823\cdot 10^{-07} & 2,411\cdot 10^{+02} & 7,32\cdot 10^{-08} & 5,09\cdot 10^{-08} & 1,13\cdot 10^{-08} \\
& 103 & 261 & 7,451\cdot 10^{-05} & 1,820\cdot 10^{+04} & 2,85\cdot 10^{-11} & 1,67\cdot 10^{-11} & 1,81\cdot 10^{-12} \\ \hline
Rf & 104 & 258 & 4,436\cdot 10^{-10} & 3,954\cdot 10^{+01 }& 1,67\cdot 10^{-13} & 8,74\cdot 10^{-14} & 5,91\cdot 10^{-15} \\
& 104 & 259 & 1,331\cdot 10^{-06 }& 3,267\cdot 10^{+02} & 2,13\cdot 10^{-16} & 9,65\cdot 10^{-17} & 3,54\cdot 10^{-18} \\
& 104 & 260 & 6,338\cdot 10^{-10} & 5,648\cdot 10^{+01} & 1,12\cdot 10^{-09} & 7,16\cdot 10^{-10} & 1,08\cdot 10^{-10} \\
& 104 & 261 & 2,091\cdot 10^{-05} & 5,134\cdot 10^{+03} & 8,73\cdot 10^{-17} & 3,87\cdot 10^{-17} & 1,31\cdot 10^{-18} \\
& 104 & 262 & 6,654\cdot 10^{-08} & 5,931\cdot 10^{+03} & 6,72\cdot 10^{-15} & 3,27\cdot 10^{-15} & 1,65\cdot 10^{-16} \\ \hline
Db & 105 & 255 & 2,535\cdot 10^{-07} & 6,223\cdot 10^{+01} & 2,29\cdot 10^{+11} & 4,08\cdot 10^{+11} & 4,48\cdot 10^{+12} \\
& 105 & 257 & 2,535\cdot 10^{-07} & 6,223\cdot 10^{+01} & 6,87\cdot 10^{+20} & 1,97\cdot 10^{+21} & 1,59\cdot 10^{+23} \\
& 105 & 261 & 3,169\cdot 10^{-07} & 7,779\cdot 10^{+01} & 3,00\cdot 10^{-34} & 5,48\cdot 10^{-35} & 4,67\cdot 10^{-38} \\
& 105 & 263 & 1,521\cdot 10^{-06} & 3,734\cdot 10^{+02} & 5,81\cdot 10^{-21} & 2,08\cdot 10^{-21} & 2,92\cdot 10^{-23} \\ \hline
Sg & 106 & 258 & 9,190\cdot 10^{-11} & 8,190\cdot 10^{+00} & 1,04\cdot 10^{+04} & 1,27\cdot 10^{+04} & 2,97\cdot 10^{+04} \\
& 106 & 259 & 7,605\cdot 10^{-08} & 1,867\cdot 10^{+01} & 4,36\cdot 10^{+06} & 6,10\cdot 10^{+06} & 2,47\cdot 10^{+07} \\
& 106 & 260 & 2,218\cdot 10^{-10} & 1,977\cdot 10^{+01} & 2,12\cdot 10^{+08} & 3,24\cdot 10^{+08} & 1,87\cdot 10^{+09} \\
& 106 & 261 & 8,239\cdot 10^{-08} & 2,022\cdot 10^{+01} & 7,76\cdot 10^{+10} & 1,34\cdot 10^{+11} & 1,33\cdot 10^{+12} \\
& 106 & 263 & 8,556\cdot 10^{-08} & 2,100\cdot 10^{+01} & 1,31\cdot 10^{+27} & 5,17\cdot 10^{+27} & 1,57\cdot 10^{+30} \\ \hline
Bh & 107 & 261 & 3,803\cdot 10^{-09} & 9,334\cdot 10^{-01} & 7,93\cdot 10^{-01} & 7,89\cdot 10^{-01} & 7,72\cdot 10^{-01} \\ \hline
Hs & 108 & 264 & 6,338\cdot 10^{-11} & 5,684\cdot 10^{+00} & 1,08\cdot 10^{+02} & 1,20\cdot 10^{+02} & 1,84\cdot 10^{+02} \\ \hline
\end{array}
$
\newpage
\noindent Table 2: Definitions and characteristic features of ($SF$, $\pi F$,
$T$)-nuclides.

\hspace*{-17mm}\begin{tabular}{|l|l|l|l|l|l|} \hline
Type of  & Pionic  &  Yield Mass  & Half Lives & Half Lives &  Dominant
  \\
nuclide&Yield&Distribution (MD)&$Z^{2}/A \leq 42.5$&$Z^{2}/A\geq  42.5$ & Unitarity\\ \hline
SF-nuclide & $\frac{\Gamma _{\pi ^{0,\pm }}}{\Gamma _{SF}}<1$ & Asymmetric MD
& $T_{SF}>T_{SF}^{C}$ & $T_{SF}<T_{SF}^{C}$ & $D_{SF}$ \\
T-nuclide & $\frac{\Gamma _{\pi ^{0,\pm }}}{\Gamma _{SF}}=1$ & Bimodal MD & $%
T_{SF}=T_{SF}^{C}$ & $T_{SF}=T_{SF}^{C}$ & $D_{\pi F}+D_{SF}$ \\
$\pi F$-nuclide & $\frac{\Gamma _{\pi ^{0,\pm }}}{\Gamma _{SF}}>1$ &
Symmetric MD & $T_{SF}<T_{SF}^{C}$ & $T_{SF}>T_{SF}^{C}$ & a new phase \\
&&&&& $D_{\pi F}-n\rightarrow p$\\ \hline
\end{tabular}
\pagebreak
% \enlargethispage{5cm} \thispagestyle{empty}
\vspace{5cm}
\begin{figure}[tbp]
\begin{center}
\hspace*{-1cm} \includegraphics[height=7cm]{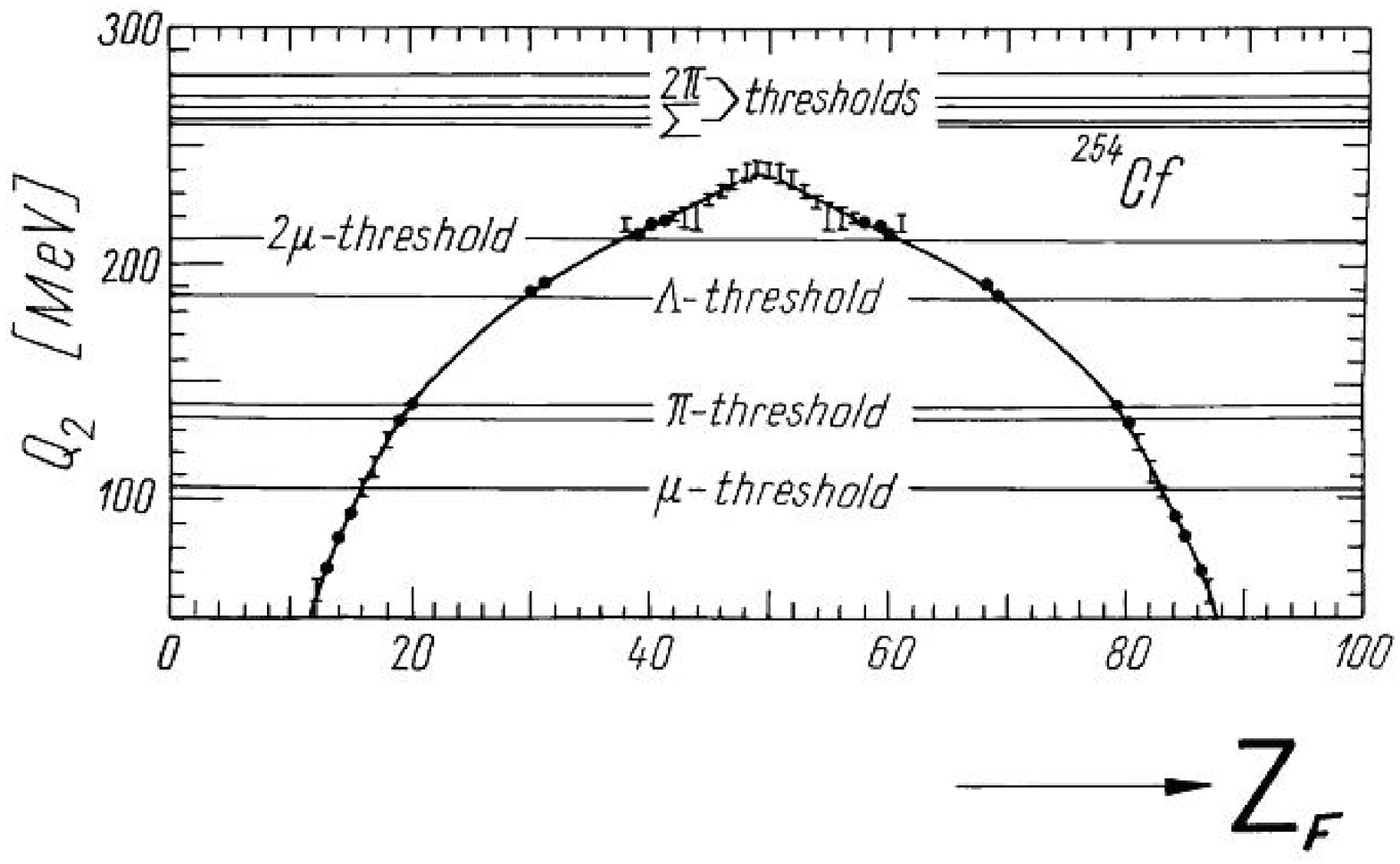}
\end{center}
\caption{$Q_2$-energy liberated in n-body spontaneous fission of some heavy
nuclei.}
\label{fg1}
\end{figure}
\nopagebreak
\begin{figure}[tbp]
\begin{center}
\hspace{-2cm} \includegraphics*[height=14cm]{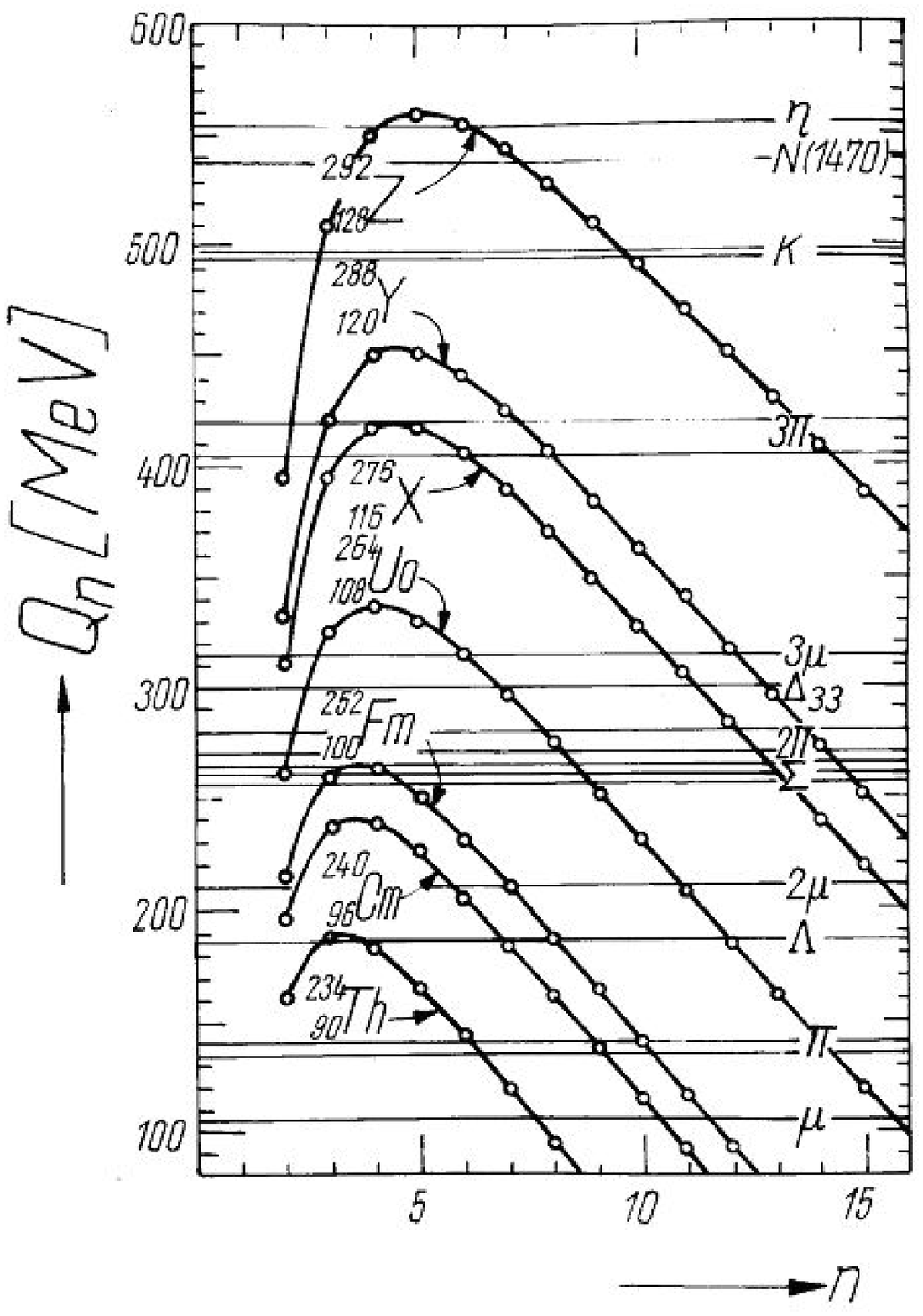}
\end{center}
\caption{$Q_n$-energy liberated in n-body spontaneous fission of some heavy
nuclei.}
\label{fg2}
\end{figure}

\newpage
\begin{figure}[tbp]
\hspace*{-25mm} \includegraphics[width=8cm]{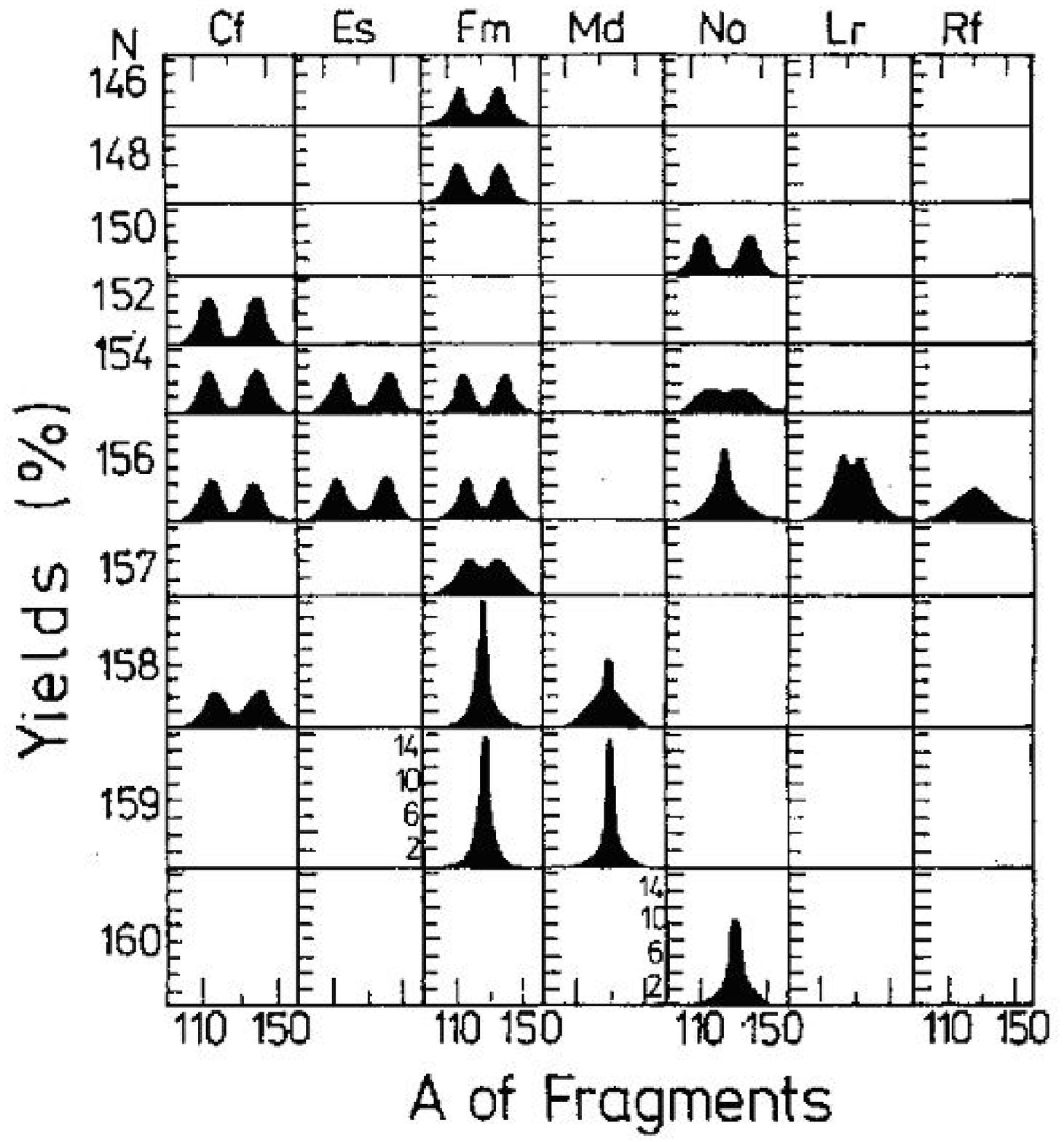} \hspace{5mm} %
\includegraphics[width=7.5cm]{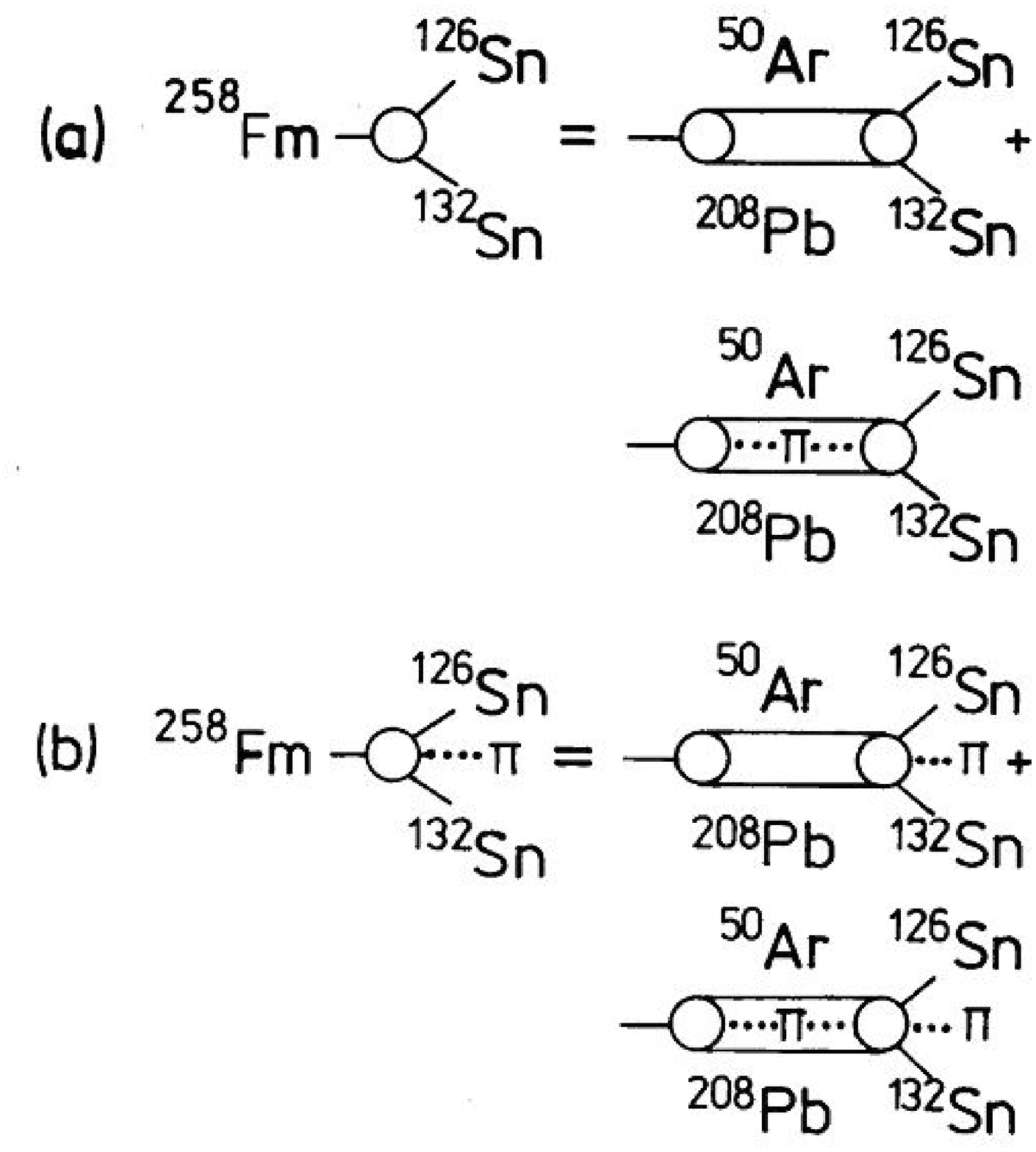}
\hspace*{-20mm} \parbox{75mm}{
 \caption{ Mass distributions of nuclear fragments for
superheavy (see Ref. [5]).}}
\hspace{5mm}
\parbox[top]{70mm}{
\caption{Description of spontaneous fission (SF)
of $^{258}Fm$ as well as of spontaneous pionic fission ($\pi F$) of $^{258}Fm$ in
terms of two types of unitarity diagrams (see the text).}}
\end{figure}

\end{document}